\documentclass[aps,eqsecnum,nofootinbib,superscriptaddress,showpacs]
{revtex4}
\usepackage{amsmath}
\newcommand{\half}{ \frac{1}{2} }
\usepackage{bm}
\bmdefine{\bmR}{ \bm{R} }
\newcommand{\hp}{ \Hat{p} }
\newcommand{\hx}{ \Hat{x} }
\newcommand{\hz}{ \Hat{z} }
\newcommand{\hU}{ \Hat{U} }
\newcommand{\vecx}{ \Vec{x} }
\newcommand{\calO}{ \mathcal{O} }
\newcommand{\jSW}{ j_{\text{SW}} }
\newcommand{\ASW}{ \mathcal{A} }
\newcommand{\BSW}{ \mathcal{B} }
\newcommand{\DSW}{ \mathcal{D} }
\newcommand{\ESW}{ \mathcal{E} }
\newcommand{\FSW}{ \mathcal{F} }
\newcommand{\JSW}{ \mathcal{J} }
\newcommand{\QSW}{ \mathcal{Q} }
\newcommand{\PhiSW}{ \Phi_{\text{SW}} }
\newcommand{\bra}[2]{\left\langle~#1~\right|}
\newcommand{\ket}[1]{\left|~#1~\right\rangle}
\newcommand{\IP}[2]{\langle~#1~\vert~#2~\rangle}
\begin{document}
\title{Noncommutative Quantum Mechanics and Seiberg-Witten Map}
\author{Akira Kokado}
\email{kokado@kobe-kiu.ac.jp}
\affiliation{Kobe International University, Kobe 658-0032, Japan}
\author{Takashi Okamura}
\email{okamura@kcs.kwansei.ac.jp}
\affiliation{Department of Physics, Kwansei Gakuin University,
Sanda 669-1337, Japan}
\author{Takesi Saito}
\email{tsaito@k7.dion.ne.jp}
\affiliation{Department of Physics, Kwansei Gakuin University,
Sanda 669-1337, Japan}
\date{\today}

\begin{abstract}
In order to overcome ambiguity problem
on identification of mathematical objects in noncommutative theory
with physical observables,
quantum mechanical system coupled to the NC U(1) gauge field
in the noncommutative space is reformulated
by making use of the unitarized Seiberg-Witten map,
and applied to the Aharonov-Bohm and Hall effects
of the NC U(1) gauge field.
Retaining terms only up to linear order in the NC parameter $\theta$,
we find that the AB topological phase and the Hall conductivity
have both the same formulas as those of the ordinary commutative space
with no $\theta$-dependence.
\end{abstract}
\pacs{11.10.Nx, 03.65.-w, 03.65.Vf}
\maketitle
\section{Introduction}\label{sec:intro}
Recently, remotivated by string theories noncommutative (NC) spacetimes
have been drawn much attention in field theories%
\cite{ref:SeibergWitten-NC,ref:Filk,ref:Schomerus,ref:MinwallaEtAL,
ref:Hayakawa,ref:MatusisEtAL,ref:DouglasNekrasov}
as well as their phenomenological implications%
\cite{ref:MocioiuEtAL,ref:CarrollEtAL,ref:HewettEtAL,ref:CarlsonEtALI,
ref:AnisimovEtAL,ref:CarlsonEtALII}.
One of the most interesting things in NC field theories is that
even the U(1) gauge group has non-Abelian like characters
such as self-interactions.

As applications of this NC U(1) gauge theory,
there are many papers concerning the Aharonov-Bohm (AB) effect%
\cite{ref:ChaichianEtAL-ABI,ref:ChaichianEtAL-ABII,ref:FalomirEtAL}
and the Hall effect%
\cite{ref:DuvalHorvathy,ref:DayiJellal,ref:KokadoEtALI,
ref:KokadoEtALII,ref:ChakraEtAL}
in the two-dimensional NC space.
However, results seem to be divergent, some show deviations%
\cite{ref:ChaichianEtAL-ABI,ref:ChaichianEtAL-ABII,ref:FalomirEtAL,
ref:DayiJellal,ref:KokadoEtALI}
and others no deviations from the ordinary commutative theories%
\cite{ref:DuvalHorvathy,ref:KokadoEtALII}.
This may come from the fact that though they discussed both effects
based on NC quantum mechanics,
but the NC U(1) gauge invariance has not been considered enough
in their papers. 

Furthermore, even if we calculate all quantities
in the NC U(1) gauge-invariant way,
we always encounter the ambiguity of how to identify
mathematical objects with physical observables%
\cite{ref:CarrollEtAL}.
For instance, the NC U(1) gauge field strength has
the NC U(1) gauge covariance, but not
the conventional U(1) gauge invariance, so that
we cannot identify the NC U(1) gauge field strength
with the physical electromagnetic field.

In order to overcome the above problem,
we make use of the Seiberg-Witten (SW) map%
\cite{ref:SeibergWitten-NC,ref:BichlEtAL}.
The SW map transforms the NC U(1) gauge  system into
the usual U(1) gauge field system in the commutative space,
and the NC U(1) gauge transformation into
the usual U(1) gauge one.
Thus, the SW map permits us to consider the NC effect
as the usual gauge theory with non-standard couplings.

In this paper, we consider the NC quantum mechanical system
minimally coupled to the NC U(1) gauge field.
By making use of the SW map, this NC system is transferred into
an equivalent commutative system with the usual U(1) gauge symmetry
but with non-minimal couplings.
As applications, we reconsider the AB effect and the Hall effect
in the NC space.

In Sec.\ref{sec:NCQM}, we reformulate NC quantum mechanics
in terms of the SW fields, which have the usual gauge symmetry
in the commutative space.
For this purpose, we use the unitarized SW map,
which is modified from the original map,
in order to make the SW map norm-preserving.

In Secs.\ref{sec:phase} and \ref{sec:Hall},
the AB topological phase and the Hall effect are considered,
respectively.
The final section is devoted to concluding remarks.
\section{Correspondence between NC quantum mechanics and
quantum mechanics in commutative space}\label{sec:NCQM}
\subsection{commutative space coordinates} 
The noncommutative space can be realized by commutation relations,
\begin{align}
  & \big[\, \hx^i\,,~\hx^j\, \big] = i\, \theta^{ij}~,
& & \big[\, \hx^i\,,~\hp_j\, \big] = i\, \delta^{i}{}_{j}~,
& & \big[\, \hp_i\,,~\hp_j\, \big] = 0~,
& & i,\, j = 1,\, 2,\, \cdots,\, n~,
\label{eq:def-NC-CCR}
\end{align}
where $\hx$ are coordinate operators,
the noncommutativity parameter, $\theta$ and
$\hp$ momentum operators.

Let us define $\hz$ by the Bopp shift
%
\cite{ref:Bopp},
\begin{equation}
  \hz^i := \hx^i + \half\, \theta^{ij}\, \hp_j~.
\label{eq:def-hz}
\end{equation}
This obeys ordinary commutation relations
\begin{align}
  & \big[\, \hz^i\,,~\hz^j\, \big] = 0~,
& & \big[\, \hz^i\,,~\hp_j\, \big] = i\, \delta^{i}{}_{j}~,
& & \big[\, \hp_i\,,~\hp_j\, \big] = 0~,
& & i,\, j = 1,\, 2,\, \cdots,\, n~.
\label{eq:def-NC-CCR-for-z}
\end{align}
Because $\hz$ is commutative,
we can define the ordinary eigenstate $\ket{\vecx}$
with eigenvalue $x^i$ of $\hz^i$ by
\begin{equation}
  \hz^i \ket{\vecx} = x^i \ket{\vecx}~.
\label{eq:eigen-eq}
\end{equation}

Since the Hamiltonian is the time-translation operator,
the Schr\"odinger equation for any state $\ket{\psi(t)}$ always holds
\begin{equation}
   H\left( \hx,\, \hp \right) \ket{\psi(t)}
  = i\, \frac{d}{dt}\, \ket{\psi(t)}~.
\label{eq:Schrodinger}
\end{equation}
By using Eq.(\ref{eq:eigen-eq}),
the coordinate representation of this equation is given by 
\begin{align}
  & \bra{\vecx} H\left( \hx,\, \hp \right) \ket{\psi(t)}
  = \bra{\vecx} H\left( \hz^j - \theta^{jk} \hp_k/2,\, \hp_j \right)
    \ket{\psi(t)}
\nonumber \\
 =&~H\left( x^j + i\,\theta^{jk} \partial_k/2,\,
           - i\, \partial_j \right) \IP{\vecx}{\psi(t)}
  = i\, \frac{\partial}{\partial t}\, \IP{\vecx}{\psi(t)}~.
\label{eq:Schrodinger-Zrep}
\end{align}
This equation (\ref{eq:Schrodinger-Zrep}) enable us
to interpret a quantum mechanics in the NC space
as the one in the commutative space.
However, there remains the problem of how to identify
the NC variables with physical observables.
\subsection{NCQM coupled to NC U(1) gauge field}
Hereafter we consider NC quantum mechanical system
minimally coupled to NC U(1) gauge filed.
The Schr\"odinger equation is given by
\begin{align}
  i \left( \frac{d}{dt} - i\, g\, A_0(\hx,\, t) \right)\, \ket{\psi(t)}
  = \frac{1}{2m}\, \big[~\hp_j - g\, A_j(\hx,\, t)~\big]^2\,
        \ket{\psi(t)}~,
\label{eq:Schrodinger-gauge}
\end{align}
where $A_\mu(\hx,\, t)$~$(\mu = 0,\, i$) is the NC U(1) gauge field.

In order to interpret the system as a usual quantum mechanics
coupled to the usual U(1) gauge field in the commutative space,
we can use the Seiberg-Witten map%
\cite{ref:SeibergWitten-NC,ref:BichlEtAL},
which is usually given as
\begin{align}
  & \ASW_\mu (\vecx,\, t)
 := A_\mu (\vecx,\, t ) + \frac{g}{2}\, A_\rho(\vecx,\, t)~
      \theta^{\rho\sigma}\, \big[\, \partial_\sigma A_\mu(\vecx,\, t)
                                  + F_{\sigma\mu}(\vecx,\, t)\, \big]~,
\label{eq:def-A-SW} \\
  & \FSW_{\mu\nu}(\vecx,\, t)
 := 2\, \partial_{[\mu} \ASW_{\nu]}(\vecx,\, t)
  = F_{\mu\nu}(\vecx,\, t) - g\, \theta^{\rho\sigma} \Big(
      F_{\mu \rho}(\vecx,\, t)\, F_{\nu \sigma}(\vecx,\, t)
    - A_\rho(\vecx,\, t) \big[~\partial_\sigma F_{\mu\nu}(\vecx,\, t)~
                         \big] \Big) + O(\theta^2)~,
\label{eq:def-F-SW} \\
  & \IP{\vecx}{ \psi_{\text{SW}}(t) }
 :=  \IP{\vecx}{\psi(t)} - \frac{ g\,\theta^{\rho\sigma} }{2}\,
          A_\sigma(\vecx,\, t)\, \partial_\rho  \IP{\vecx}{\psi(t)}~,
\label{eq:psi_SW-vs-psi-Zrep}
\end{align}
where $F_{\mu\nu}$ is the NC U(1) gauge field strength defined by
\begin{equation}
  F_{\mu\nu} := \partial_\mu A_\nu - \partial_\nu A_\mu
     - i\, g\, \big[\, A_\mu,\,~A_\nu\, \big]_\star~,
\label{eq:def-NCF}
\end{equation}
with the Moyal $\star$-product.

However the transformation (\ref{eq:psi_SW-vs-psi-Zrep})
is not unitary, that is, not norm preserving.
Since it is convenient to use unitary transformations
in (NC) quantum mechanics,
we make the usual SW map so as to be unitary in the following: 
The desired map is induced by the unitary operator,
\begin{equation}
  \hU := \exp\Big( - i\, \frac{g}{4}\, \theta^{kl}
           \big[~p_k A_l(\hx,\, t) + A_l(\hx,\, t) p_k~\big] \Big)~.
\label{eq:def-hU}
\end{equation}
This induces a transformation in a state as
\begin{align}
  & \ket{\psi(t)}~\longrightarrow~\ket{\psi'(t)}
  = \hU \ket{\psi(t)}~,
\label{eq:unitary-state} \\
  & \IP{\vecx}{ \psi' }
  =  \IP{\vecx}{ \psi } - \frac{ g\,\theta^{kl} }{2}\, \partial_k
        \Big[~A_l(\vecx,\, t)\, \IP{\vecx}{ \psi }~\Big]
  + O(\theta^2)~,
\label{eq:psi_prime-vs-psi-Zrep}
\end{align}
similar to the usual SW map (\ref{eq:psi_SW-vs-psi-Zrep}).
Under the NC U(1) gauge transformation,
$\delta \IP{\vecx}{\psi} = i\, \lambda \star \IP{\vecx}{\psi}$,
the new field $\IP{\vecx}{\psi'}$ is transformed as a fundamental field
of the usual U(1) gauge group, that is,
\begin{align}
  & \delta \IP{\vecx}{ \psi' }
  = i\, g\, \lambda_{\text{SW}}\, \IP{\vecx}{ \psi' } + O(\theta^2)~,
\label{eq:psi-prime-trans}
\end{align}
where
\begin{align}
   \lambda_{\text{SW}}
  &:= \lambda + \frac{g}{2}\, \theta^{\rho\sigma}
                               A_\rho (\partial_\sigma \lambda)~,
\label{eq:def-lambda-SW}
\end{align}
and also,
\begin{align}
  & \delta \ASW_\mu
  = \partial_\mu \lambda_{\text{SW}} + O(\theta^2)~,
& & \delta \FSW_{\mu\nu} = 0 + O(\theta^2)~.
\label{eq:calA-calF-SW-trans}
\end{align}

In terms of the commutative coordinate (\ref{eq:def-hz}), i.e.,
$\hz^i = \hx^i - \theta^{ij} \hp_j/2$ and
the Seiberg-Witten fields $\ASW$, $\FSW$ and $\ket{\psi'}$,
we obtain
\begin{align}
  & \hU\, \big(\, \hp_i - g\, A_i(\hx,\, t)\, \big)\, \hU^\dagger
  = \hp_i - g\, \ASW_i(\hz,\, t) + \frac{ g\, \theta^{kl} }{2}\,
               \Hat{\calO}_{kl,i}(\hz,\, t) + O(\theta^2)~,
\label{eq:transfer-minimal} \\
  & \hU\, i \left( \frac{d}{dt} - i\, g\, A_0(\hx,\, t) \right)
     \hU^\dagger
   = i\, \left( \frac{d}{dt} - i\, g\, \ASW_0(\hz,\, t)
       + i\, \frac{ g\, \theta^{kl} }{2}
          \Hat{\calO}_{kl,0}(\hz,\, t) \right) + O(\theta^2)~,
\label{eq:transfer-time-component}
\end{align}
where
\begin{align}
   \Hat{\calO}_{kl,\mu}(\hz,\, t)
  &:= \frac{1}{2}\, \big\{~\hp_k - g\, \ASW_k(\hz,\, t)~,~
       \FSW_{\mu\, l}(\hz,\, t)~\big\}~,
\label{eq:def-calO-mu}
\end{align}
and $\{~,~\}$ is the anti-commutator.
$\ASW(\hz,\, t)$ and $\FSW(\hz,\, t)$ are operator representations
of Eqs.(\ref{eq:def-A-SW}) and (\ref{eq:def-F-SW}),
\begin{align}
  & \ASW_\mu (\hz,\, t)
 := A_\mu (\hz,\, t ) + \frac{g}{2}\, A_\rho(\hz,\, t)~
      \theta^{\rho\sigma}\, \big[\, \partial_\sigma A_\mu(\hz,\, t)
                                  + F_{\sigma\mu}(\hz,\, t)\, \big]~,
\label{eq:def-A-SW-op} \\
  & \FSW_{\mu\nu}(\hz,\, t) := 2\, \partial_{[\mu} \ASW_{\nu]}
  = F_{\mu\nu}(\hz,\, t) - g\, \theta^{\rho\sigma} \Big(
      F_{\mu \rho}(\hz,\, t)\, F_{\nu \sigma}(\hz,\, t)
    - A_\rho(\hz,\, t) \big[~\partial_\sigma F_{\mu\nu}(\hz,\, t)~\big]
                                                  \Big) + O(\theta^2)~.
\label{eq:def-calF-SW-op}
\end{align}

Finally, by collecting Eqs.(\ref{eq:transfer-minimal})
and (\ref{eq:transfer-time-component}),
the Schr\"odinger equation for $\ket{\psi'(t)}$ becomes
\begin{align}
  & i\, \left( \frac{d}{dt} - i\, g\, \ASW_0(\hz,\, t)
     + i\, \frac{ g\, \theta^{kl} }{2}\,
          \Hat{\calO}_{kl,0}(\hz,\, t) \right) \ket{\psi'(t)}
   = \frac{1}{2m}\, \left( \hp_i - g\, \ASW_i(\hz,\, t)
       + \frac{ g\, \theta^{kl} }{2}\,
          \Hat{\calO}_{kl,i}(\hz,\, t) \right)^2 \ket{\psi'(t)}~.
\label{eq:Schrodinger-SW}
\end{align}
This equation (\ref{eq:Schrodinger-SW}) permits us to interpret
a NC quantum mechanics minimally coupled to the NC U(1) gauge field
as a quantum mechanics non-minimally coupled to
a usual U(1) gauge field in commutative space.

It is worthwhile to note that,
when we regard the SW gauge field as the observable U(1) gauge field,
NC correction terms come from $\calO_{kl,\mu}$,
which is proportional to the (observable) SW field strength $\FSW$.
Therefore, we conclude that,
in the region where the SW field strength $\FSW$ vanishes,
we cannot detect the NC effect.
\section{NC AB effect and NC Hall effect Revisited} 
\subsection{The AB topological phase} \label{sec:phase}
In the AB effect (and also Hall effect),
we are interested in a stationary configuration
$\ASW = \ASW(\vecx)$,
and look for the stationary state with energy $\omega$ in
the commutative coordinate representation.
So we can put as
\begin{equation}
  \IP{\vecx}{\psi'(t)} = \varphi(\vecx)\, \exp( - i \omega t )~,
\label{eq:ansatz}
\end{equation}
to give
\begin{align}
  & \Big( \omega + g\, \ASW_0(\vecx)
              - \frac{ g\, \theta^{kl} }{2}\,
                   \Hat{\calO}_{kl,0}(\vecx) \Big) \varphi(\vecx)
   = \frac{-1}{2m}\, \Big( \partial_i - i\, g\, \ASW_i(\vecx)
           + i\, \frac{ g\, \theta^{kl} }{2}\,
                   \Hat{\calO}_{kl,i}(\vecx) \Big)^2 \varphi(\vecx)~.
\label{eq:Schrodinger-SW-Zrep}
\end{align}

Now, let us assume $\ASW_0 = 0$.
We define the area $D$ by
\begin{equation}
   \FSW_{ij}(\vecx)
  = \begin{cases}
         \epsilon_{ijk}\, \BSW^k (\vecx)& \text{~for~~} \vecx \in D \\
                                      0 & \text{~for~~} \vecx \notin D
    \end{cases}~,
\label{eq:domain-D}
\end{equation}
where $\BSW$ is the SW magnetic field.
Since $\FSW_{ij} = 0$ outside $D$, the Schr\"odinger equation becomes
\begin{equation}
   \omega\, \varphi
  = \frac{-1}{2m}\, \left( \partial_j - i\, g\, \ASW_j \right)^2
         \varphi~.
\label{eq:Schrodinger-SW-Zrep-AB-out}
\end{equation}

Therefore, following the ordinary procedure,
we obtain the AB topological phase,
\begin{align}
  & \Theta(C)
 := g \oint_{C} dx^k~\ASW_k(\vecx)
  = g \int_{\text{int}(C)} dS_i~\BSW^i(\vecx)
  = g \int_{D} dS_i~\BSW^i(\vecx)
  = g\, \PhiSW(D)~.
\label{eq:def-Theta}
\end{align}
Due to U(1) character of the SW field for NC U(1) gauge transformation,
the magnetic flux $\PhiSW(D)$ of the SW gauge field is
NC U(1) gauge invariant.

In conclusion, the AB phase in the NC case is the same as
the ordinary AB phase in the commutative case
without $\theta$-dependent term.

We compare our result with existent results%
\cite{ref:ChaichianEtAL-ABI,ref:ChaichianEtAL-ABII,ref:FalomirEtAL},
which assert that one can detect the noncommutativity of the space.
Chaichian et al.%
\cite{ref:ChaichianEtAL-ABI,ref:ChaichianEtAL-ABII}
and Falomir et al.%
\cite{ref:FalomirEtAL}
estimated the holonomy of 
the NC U(1) gauge field $A$, and obtained the result
\begin{align}
   \Theta_{\text{exist.}}(C)
  &= - \int_{C} dx^i~\left( A_i + \half\, \theta^{jk}\,
     \big[\, m\, v_j \partial_k A_i - A_j\partial A_i\, \big] \right)~.
\label{eq:delta-phi-existence}
\end{align}
On the other hand, our holonomy in the region of $\FSW=0$
is expressed as,
\begin{align}
   \Theta_{\text{ours}}(C)
  &= (-1) \int_{C} dx^i~\ASW_i(\vecx)
   = - \int_{C} dx^i~\left( A_i + \frac{(-1)}{2}\, A_j~\theta^{jk}\,
        \big[\, \partial_k A_i + F_{ki}\, \big] \right)
\nonumber \\
  &= \Theta_{\text{exist.}}(C)
   + \frac{m}{2}\, \theta^{jk} \int_{C} dx^i~v_j \partial_k A_i
   + \frac{\theta^{jk}}{2}\, \int_{C} dx^i~A_j\, F_{ki}~,
\label{eq:delta-phi-ours}
\end{align}
where we take $g=-1$ in order to compare with theirs
\footnote{They use the covariant derivative,
$D_i = \partial_i + i\, A_i$.}.

The second term in Eq.(\ref{eq:delta-phi-ours}) can be rewritten as
\begin{align}
   \theta^{jk} \int_{C} dx^i~v_j \partial_k A_i
  &= \theta^{jk} \int dt~\frac{dx^i}{dt}\, \frac{dx_j}{dt}\,
        \partial_k \ASW + O(\theta^2)
\nonumber \\
  &= \theta^{jk} \int dt~\frac{dx^i}{dt}\, \frac{dx_j}{dt}\,
             \FSW_{ki}
   - \theta^{jk} \int dt~\frac{d^2x_j}{dt^2}\, \ASW_k
   + \theta^{jk} \int dt~\frac{d}{dt}
             \left( \frac{dx_j}{dt}\, \ASW_k \right)
   + O(\theta^2)~.
\label{eq:delta-phi-difference}
\end{align}
Because they estimated the holonomy 
$\Theta_{\text{exist.}}$ under the semiclassical approximation,
we can use the equation of motion to estimate the second term
in Eq.(\ref{eq:delta-phi-difference}).
The equation of motion at the zeroth order in $\theta$ is given by
\begin{equation}
  m\, \frac{d^2 x_j}{dt^2}
  = g\, \FSW_{jk} \frac{dx^k}{dt} = - \FSW_{jk} \frac{dx^k}{dt}~.
\label{eq:EOM}
\end{equation}
Finally, we obtain
\begin{align}
   \Theta_{\text{ours}}(C)
  &= \Theta_{\text{exist.}}(C)
   + \frac{m}{2}\, \theta^{jk}\int_{C} dx^i~\frac{dx_j}{dt}\, \FSW_{ki}
   + \frac{m}{2}\, \theta^{jk} \int_{C} dx^i~
        \partial_i \left( \frac{dx_j}{dt}\, \ASW_k \right)
   + O(\theta^2)~.
\label{eq:delta-phi-oursII}
\end{align}
In the region of $\FSW = 0$, the second term
in Eq.(\ref{eq:delta-phi-oursII}) vanishes.
The third term is a surface term, so that for the closed contour
it vanishes,
if $\dot{x}_j A_k$ is a one-valued function.

Thus, we have $\Theta_{\text{exist.}}(C) = \Theta_{\text{ours}}(C)$,
that is, their results for a closed contour outside the solenoid
($\FSW = 0$) are rewritten in terms of the holonomy
of the SW gauge field.
So we cannot detect the noncommutativity of the space
through the AB effect, if we identify the SW gauge field strength
as the physical one.
\subsection{The Hall effect in the NC space} \label{sec:Hall}
In this section, we reconsider the Hall effect in the NC space
by using the unitarized Seiberg-Witten map.
As is mentioned in the introduction,
even in the \lq\lq classical\rq\rq\, Hall effect without
impurity and many-body interaction in the NC space,
there are divergent results.
Therefore, we confine ourselves into
the \lq\lq classical\rq\rq\, Hall effect and do not consider
the integer/fractional quantum Hall effect.

We start with the general formula (\ref{eq:Schrodinger-SW-Zrep}).
By taking the Landau gauge%
\footnote{In this case, using Eq.(\ref{eq:def-F-SW}),
the NC U(1) magnetic field $B^3 := F_{12}$ is related to
the SW magnetic field $\BSW$ as
$B^3 = \BSW ( 1 + g\, \theta \BSW ) + O(\theta^2)
= \BSW/( 1 - g\, \theta \BSW ) + O(\theta^2)$.
Thus, the SW magnetic field $\BSW$ coincides with
Nair and Polychronakos' U(1) gauge magnetic field%
\cite{ref:NairPoly},
up to $O(\theta)$.
}
\begin{align}
  & \ASW_0( \vecx,\, t ) = \ESW\, x^1~,
& & \ASW_1( \vecx,\, t ) = 0~,
& & \ASW_2( \vecx,\, t ) = \BSW\, x^1~,
\label{eq:Landau-gauge}
\end{align}
Eq.(\ref{eq:Schrodinger-SW-Zrep}) reduces to
\begin{align}
  & \left( \omega + g\, \ESW\, x^1 + i\, \frac{g\, \theta}{2}\, \ESW
       ( \partial_2 - i\, g\, \BSW x^1 ) \right) \varphi(\vecx)
  = \frac{-1}{2m}\, ( 1 + g\, \theta\, \BSW )
       \Big[~\partial_1^2 + ( \partial_2 - i\, g\, \BSW x^1 )^2~\Big]
          \varphi(\vecx)~.
\label{eq:Schrodinger-SW-Zrep-Landau}
\end{align}

Separating variables as
$\varphi(\vecx) = \phi(x^1) \exp( i\, p_2\, x^2 )$, we get
\begin{align}
   E\, \phi(X)
  &= \left( \frac{-1}{2m}\, \frac{\partial^2}{\partial X^2}
       + \frac{m \omega_c^2}{2}\, X^2 \right) \phi(X)~,
\label{eq:Schrodinger-X}
\end{align}
where 
\begin{align}
  & E := \left( \omega + \frac{p_2 \ESW}{ \BSW }
    + \frac{m}{2}\, \frac{ \ESW^2 }{ \BSW^2 } \right)
      \left( 1 - g\, \theta\, \BSW \right)~,
& & X := x^1 - \left(
         \frac{p_2}{ g \BSW } + \frac{ m \ESW }{ g\, \BSW^2 }\,
                    \left( 1 - \frac{g\,\theta}{2}\, \BSW \right)
               \right)~,
\label{eq:def-X}
\end{align}
and $\omega_c := g\, \BSW/m$.
The eigenvalue and eigenstate of Eq.(\ref{eq:Schrodinger-X})
are well known to be
\begin{align}
  & E_n = \left( n + \half \right) \omega_c~,
& & \phi_n(X) = C_n\, \exp\left( - \frac{m\, \omega_c}{2}\, X^2 \right)
      H_n\big( \sqrt{ m\, \omega_c }~X \big)~,
\label{eq:E-eigen-state}
\end{align}
where $C_n$ is a normalization constant and
$H_n$ is the $n$-th Hermite polynomial.

Now we consider the current coupled to the SW U(1) gauge field $\ASW$.
From the action,
\begin{align}
   S
  &= \int dt d^2x~\varphi^* \left[~i \left( \partial_t - i\, g\, \ASW_0
      + i\, \frac{ g\, \theta^{kl} }{2}\, \Hat{\calO}_{kl,0} \right)
     + \frac{1}{2m}\, \left( \partial_i - i\, g\, \ASW_i
      + i\, \frac{ g\, \theta^{kl} }{2}\, \Hat{\calO}_{kl,i} \right)^2~
                            \right] \varphi~,
\label{eq:def-action}
\end{align}
the expectation value of the matter current density coupled to $\ASW$
is given by
\begin{align}
  & \jSW^\mu( \vecx,\, t )
 := \frac{ \delta S }{ \delta \ASW_\mu( \vecx,\, t ) }~.
\label{eq:def-jSW}
\end{align}
In the gauge configuration (\ref{eq:Landau-gauge}),
the expectation values of the total charge and total currents,
$\displaystyle{ \JSW^{\mu} := \int d^2x~\jSW^{\mu}( \vecx,\, t ) }$,
become (see Appendix \ref{sec:append})
\begin{align}
   \QSW &:= \JSW^0
  = g\, \int d^2x~\vert\, \varphi_n\, \vert^2 = g~,
\label{eq:JSW-zero} \\
   \JSW^2
  &= \frac{g}{m}\, \left( 1 + g\, \theta\, \BSW \right)
       \int d^2x~\Im\big[\, \varphi^* ( \DSW_2 \varphi )\, \big]
   + \frac{g^2\, \theta}{2}\, \ESW
        \int d^2x~\vert\, \varphi_n\, \vert^2
   = - \frac{g\, \ESW}{\BSW}\,
            \left( 1 + \frac{g\, \theta}{2}\, \BSW \right)
   + \frac{g^2\, \theta}{2}\, \ESW
   = - \frac{g\, \ESW}{\BSW}~,
\label{eq:JSW-two} \\
   \JSW^1 &= 0~,
\label{eq:JSW-one}
\end{align}
where we have used the equation
\begin{align}
   \int d^2x~\Im\big[\, \varphi^* ( \DSW_2 \varphi )\, \big]
  &= \int d^2x~\left( p_2 - g\, \BSW\, x^1 \right)
          \vert\, \varphi_n\, \vert^2
   = \int d^2x~\left[~- g\, \BSW\, X
        - \frac{m\, \ESW}{\BSW}\,
            \left( 1 - \frac{g\, \theta}{2}\, \BSW \right)\, \right]
          \vert\, \varphi_n\, \vert^2
\nonumber \\
  &= - \frac{m\, \ESW}{\BSW}\,
            \left( 1 - \frac{g\, \theta}{2}\, \BSW \right)~.
\label{eq:pre-JSW-two}
\end{align}

From Eq.(\ref{eq:JSW-two}), we have the Hall conductivity,
which coincides with the ordinary Hall conductivity,
\begin{equation}
   \sigma_{\text{NC}} = - g/\BSW = - \QSW/\BSW~.
\label{eq:sigma_H-SW}
\end{equation}
%
\section{Concluding remarks}\label{sec:conclude}
We have considered the NC quantum mechanical system
minimally coupled to the NC U(1) gauge field.
By making use of the unitarized SW map,
this NC system has been transferred into
an equivalent commutative system with the usual U(1) gauge symmetry
but with non-minimal couplings.
As applications, we have reconsidered the AB effect and the Hall effect
in the NC space.
The AB topological phase is just the SW magnetic flux (times $g$)
through the domain $D$.
Thus, the AB phase in the NC space is the same
as the ordinary AB phase in the commutative space
without $\theta$-dependent term.
The same thing also happens on the Hall conductivity
in the NC space, which is given by
$\sigma_{\text{NC}} = - g/\BSW = - \QSW/\BSW$.
%

We note that
the SW magnetic flux $\PhiSW$ is also equal to
the \lq\lq NC magnetic flux\rq\rq~in the case of
a homogeneous configuration along the NC magnetic field line.
This can be seen as follows:
Let $\Sigma$ denote the two-dimensional plane
normal to the NC magnetic field line and
choose $\vecx_\perp = ( x^1,\, x^2 )$
as the two-dimensional coordinates on $\Sigma$.
We also use $x^3$ as the coordinate along the NC magnetic field line.
Furthermore, let $D$ denote the support of the NC magnetic field
on $\Sigma$.
Although we cannot generally define the \lq\lq NC magnetic flux\rq\rq~
in a gauge-invariant manner,
in the case of the homogeneous configuration,
we have a natural gauge invariant \lq\lq NC magnetic flux\rq\rq~as,
\begin{align}
   \Phi_{\text{NC}}
 &:= \int_{\bmR^3} d^3 x~B^3(\vecx_\perp) \Big/ \int d(x^3)
  = \int_{\bmR^3} d(x^3) d^2 x_\perp~B^3(\vecx_\perp) \Big/ \int d(x^3)
  = \int_{\bmR^2} d^2 x_\perp~B^3(\vecx_\perp)~.
\label{eq:def-Phi_NC}
\end{align}
Because of the equality
\begin{align}
   \int_{\bmR^3} d^3x~U \star F_{\mu\nu} \star U^\dagger
  = \int_{\bmR^3} d^3x~F_{\mu\nu} \star U^\dagger \star U
  = \int_{\bmR^3} d^3x~F_{\mu\nu}~,
\end{align}
the $\Phi_{\text{NC}}$ is gauge invariant.

For the configuration,
$\BSW^1 = \BSW^2 = 0$ and $\BSW^3( \vecx_\perp )$ with the support $D$,
we obtain $B^1 = B^2 = 0$ and $B^3( \vecx_\perp )$ in the gauge,
$\ASW_1(\vecx_\perp)$, $\ASW_2(\vecx_\perp)$ and $\ASW_3 = 0$,
by using the relation between the SW magnetic field $\BSW$ and
the NC magnetic field $B^i := (1/2) \epsilon^{ijk} F_{jk}$,
\begin{align}
  & B^i = \BSW^i + \partial_j \left(
          g\, \theta^{jk} \ASW_k \BSW^i \right) + O(\theta^2)~.
\label{eq:BSW-B}
\end{align}
We obtain
\begin{align}
   \PhiSW(D)
  &= \int_{D} d^2x_\perp~\BSW^3(\vecx_\perp)
   = \int_{\bmR^2} d^2x_\perp~\BSW^3(\vecx_\perp)
   = \int_{\bmR^2} d^2x_\perp~\left[~B^3
      - \partial_j \left( g\, \theta^{jk} \ASW_k \BSW^3 \right)~\right]
   + O(\theta^2)
\nonumber \\
  &= \int_{\bmR^2} dx^1 dx^2~\left[~B^3
      - \partial_1 \left( g\, \theta^{12} \ASW_2 \BSW^3 \right)
      - \partial_2 \left( g\, \theta^{21} \ASW_1 \BSW^3 \right)~\right]
   + O(\theta^2)
\nonumber \\
  &= \int_{\bmR^2} dx^1 dx^2~B^3 = \Phi_{\text{NC}}~,
\label{eq:relation-Phi}
\end{align}
because of $\BSW^3 = 0$ outside the domain $D$.

In any case, we cannot detect the NC effect
through the AB effect and the (classic) Hall effect.
However, in any real experiment of the Hall effect,
we must incorporate the impurity effect and many body correlation
among the charged particles,
so that it is very interesting to consider
the integer/fractional Hall effect in the NC space.
\appendix
\section{the current density and the total current}
\label{sec:append}
The action (\ref{eq:def-action}) of the field $\varphi$ is rewritten by
\begin{align}
   S[\, \varphi,\, \ASW\, ]
  &= \int dt\, d^{n-1} x~\left[~\varphi^* i\, \partial_0 \varphi
       + g\, \ASW_0\, \vert \varphi \vert^2
       + \frac{g\, \theta^{kl}}{2}\, \FSW_{0l}~
            \Im\big[\, ( \DSW_k \varphi )^*\, \varphi\, \big] \right.
\nonumber \\
  &\hspace*{1.5cm} \left.
   + \frac{-1}{2m}\, \left\{ \left( \delta^{kl}
         - g\, \theta^{m(k} \FSW^{l)}{}_m \right)
             ( \DSW_k \varphi )^* ( \DSW_l \varphi )
       + \frac{ g\, \theta^{kl} }{4}\, ( \partial_k \FSW^i{}_l )~
          \partial_i \vert\, \varphi\, \vert^2 \right\}~
   \right] + O(\theta^2)
   =: \int d^{n}x~\mathcal{L}~,
\label{eq:action}
\end{align}
where $\DSW_\mu$ is the covariant derivative with respect to
the SW U(1) gauge field $\ASW_\mu$,
$\DSW_\mu := \partial_\mu - i\, g\, \ASW_\mu$.

The expectation value of the matter current density coupled to $\ASW$
defined by Eq.(\ref{eq:def-jSW}) becomes
\begin{align}
  & \jSW^0( \vecx,\, t )
  = g\, \vert \varphi \vert^2 - \frac{ g\, \theta^{kl} }{2}\,
         \partial_l~\Im\big[\, \varphi^* ( \DSW_k \varphi )\, \big]~,
\label{eq:jSW_zero} \\
  & \jSW^i( \vecx,\, t )
  = \frac{g}{m}\, \left\{ \delta^{ij}
      - g\, \theta^{k(i} \FSW^{j)}{}_k \right\}
         \Im\big[\, \varphi^* ( \DSW_j \varphi )\, \big] 
  + \frac{g}{m}\, \partial_j~\Re\big[\, ( \partial_k \varphi )^*
                            \theta^{k[i} ( \DSW^{j]} \varphi )\, \big]
  + \frac{ g\, \theta^{ij} }{4m}\, \partial_j \triangle
       \vert \varphi \vert^2
\nonumber \\
  & \hspace*{2cm}
  - \frac{ g\, \theta^{ij} }{2}\, \partial_0~
       \Im\big[\, \varphi^* ( \DSW_j \varphi )\, \big] 
  + \frac{ g^2\, \theta^{ij} }{2}\, \FSW_{0j}~\vert \varphi \vert^2~,
\label{eq:jSW_spatial}
\end{align}
where we have used equations,
\begin{align}
  & \frac{ \delta ( \DSW_\nu \varphi )(y) }{ \delta \ASW_\mu(x) }
  = - i\, g\, \delta^\mu_\nu\, \delta(x-y)\, \varphi(y)~,
& & \frac{ \delta \FSW_{\rho\sigma}(y) }{ \delta \ASW_\mu(x) }
  = 2\, \delta^\mu_{[\sigma}\, \partial^y_{\rho]}\, \delta(x-y)
  = 2\, \delta^\mu_{[\rho}\, \partial^x_{\sigma]}\, \delta(x-y)~.
\label{eq:formulaII-append}
\end{align}
Therefore, the expectation value of the total current,
$\displaystyle{ \JSW^{\mu} := \int d^2x~\jSW^{\mu}( \vecx,\, t ) }$,
becomes
\begin{align}
  & \QSW := \JSW^0
  = g\, \int d^{n-1}x~\vert\, \varphi_n\, \vert^2 = g~,
\label{eq:JSW-time} \\
  & \JSW^i
  = \int d^{n-1}x~ \left[~\frac{g}{m}\, \left\{ \delta^{ij}
                         - g\, \theta^{k(i} \FSW^{j)}{}_k \right\}
       \Im\big[\, \varphi^* ( \DSW_j \varphi )\, \big] 
  - \frac{ g\, \theta^{ij} }{2}\, \partial_0~
       \Im\big[\, \varphi^* ( \DSW_j \varphi )\, \big] 
  + \frac{ g^2\, \theta^{ij} }{2}\, \FSW_{0j}~
         \vert \varphi \vert^2~\right]~.
\label{eq:JSW-spatial}
\end{align}

For a stationary configuration, we have a convenient formula
for the expectation value of the total current, as follows:
By the spatially homogeneous variation,
$\ASW_\mu(\vecx)~\mapsto~\ASW_\mu(\vecx) + a_\mu$,
where $a_\mu$ is a constant vector,
the change in the action up to $O(a)$ is given by
\begin{align*}
  & \delta S
  = \int d^n x~\frac{\delta S}{\delta \ASW_\mu(\vecx,\, t)}\, a_\mu
  = a_\mu \int d^n x~\frac{\delta S}{\delta \ASW_\mu(\vecx,\, t)}
& & \longrightarrow
& & \left. \frac{ \partial S[\, \varphi,\, \ASW + a\, ] }
                { \partial a_\mu }\, \right\vert_{a=0}
  = \int d^n x~\frac{\delta S}{\delta \ASW_\mu(\vecx,\, t)}
  = \int dt~\JSW^\mu(t)~,
\end{align*}
so that, for a stationary configuration, we obtain a convenient formula
%
\begin{align}
  & \JSW^\mu
  = \left. \frac{ \partial S }{ \partial a_\mu }\, \right\vert_{a=0}
    \bigg/ \int dt
  = \int d^{n-1}x~\frac{ \partial \mathcal{L} }{ \partial \ASW_\mu }~.
\label{eq:JSW-simple}
\end{align}
%

\end{document}